\documentclass[aps,pra,amsmath,amssymb,reprint]{revtex4-2}

\usepackage{graphicx}
\usepackage{dcolumn}
\usepackage{bm}
\usepackage{multirow}
\usepackage{tikz}
\usepackage{comment}

\newcommand{\subfigimg}[3][,]{%
  \setbox1=\hbox{\includegraphics[#1]{#3}}
  \leavevmode\rlap{\usebox1}
  \rlap{\hspace*{-8pt}\raisebox{\dimexpr\ht1-2\baselineskip/2}{#2}}
  \phantom{\usebox1}
}

\graphicspath{ {./Figures/} }
\begin{document}


\title{Simulation Method for Investigating the Use of Transition-Edge Sensors as Spectroscopic Electron Detectors}

\author{K. M. Patel}
\author{S. Withington}%
\author{C. N. Thomas}%
\author{D. J. Goldie}%
\affiliation{Cavendish Laboratory, University of Cambridge, 
JJ Thomson Avenue, Cambridge CB3 0HE, United Kingdom}%

\author{A. G. Shard}
\affiliation{%
National Physical Laboratory, Hampton Road, Teddington TW11 0LW, United Kingdom}%

\date{\today}

\begin{abstract}

Transition-edge sensors (TESs) are capable of highly accurate single particle energy measurement. TESs have been used for a wide range of photon detection applications, particularly in astronomy, but very little consideration has been given to their capabilities as electron calorimeters. Existing electron spectrometers require electron filtering optics to achieve energy discrimination, but this step discards the vast majority of electrons entering the instrument. TESs require no such energy filtering, meaning they could provide orders of magnitude improvement in measurement rate. To investigate the capabilities of TESs in electron spectroscopy, a simulation pipeline has been devised. The pipeline allows the results of a simulated experiment to be compared with the actual spectrum of the incident beam, thereby allowing measurement accuracy and efficiency to be studied. Using Fisher information, the energy resolution of the simulated detectors was also calculated, allowing the intrinsic limitations of the detector to be separated from the specific data analysis method used. The simulation platform has been used to compare the performance of TESs with existing X-ray photoelectron spectroscopy (XPS) analysers.  TESs cannot match the energy resolution of XPS analysers for high-precision measurements but have comparable or better resolutions for high count rate applications. The measurement rate of a typical XPS analyser can be matched by an array of 10 TESs with 120\,$\mu s$ response times and there is significant scope for improvement, without compromising energy resolution, by increasing array size.
\end{abstract}

\maketitle

\section{\label{sec:level1}Introduction}

The most commonly used electron detectors determine particle flux by measuring the flow of charge, either directly or after charge multiplication. Existing detectors have little or no inherent energy resolution, and so energy measurement must be achieved by rejecting electrons that fall outside of some specified energy band. By sweeping the band across the spectral range of interest, a complete spectrum can be assembled. Because the energy resolution is determined by the width of the filter, the higher the spectral resolution, the smaller the measurement rate, and this trade-off is a significant problem in many applications. Using X-ray photoelectron spectroscopy (XPS) as an example, an analyser's resolution can be better than 0.1\,eV full-width half maximum when measuring a narrow portion of a spectrum, but must be increased to several electron-volts for wide spectral scans (see Section 4). The energy filtering process discards the vast majority of electrons available to the instrument ($>$99\% at any one time) severely limiting the available count rate, and making high-resolution, full-spectrum measurements prohibitively time-consuming. Improvements can be made to the total number of electrons entering an instrument by widening the collection angle, or increasing the electron-source intensity, but these improvements only partially mitigate the innate inefficiency of the measurement technique. A detector capable of measuring electron energies without prior energy filtering would be able to provide orders of magnitude improvement in measurement efficiency.

Transition-edge sensors (TESs) are highly-sensitive radiation and particle detectors, which in their calorimetric mode of operation, do not require energy filtering to achieve energy-resolved measurements. As such, they are a potential candidate for a range of applications, including XPS.
Large arrays of TESs can be fabricated lithographically and operated simultaneously, increasing the available measurement rate without compromising energy resolution; multiplexed arrays numbering thousands of TESs are currently being developed for space-based astronomical X-ray measurements \cite{taralli2021performance,smith2020toward}. TESs are also being developed for infrared, optical and UV spectroscopy in the context of emerging quantum experiments and technologies \cite{ahmed2018quantum}.

Little attention has been paid to using TESs as massive particle spectrometers. The HOLMES \cite{HOLMES} and PTOLEMY \cite{PTOLEMY} projects are both developing TESs that will measure electron energies, but as a means of determining the mass of the neutrino and measuring the cosmic neutrino background respectively. HOLMES aims to measure electron emission from the decay of holmium atoms embedded inside the sensor itself, whereas PTOLEMY will measure electrons with electron-volt energies, far lower than the energies found in electron spectroscopy. For both of these projects, the detector designs, experimental configurations, and targeted measurement rates are entirely different to those that are required for surface-science applications of electron spectroscopy.

To the authors' knowledge, the use of TESs for direct electron spectroscopy over the range 200--2000\,eV has not been investigated previously. To study their potential, to analyse possible modes of operation, and to identify the limits of likely performance, a simulation and analysis pipeline has been devised. Simulation software was written in MATLAB 2020a \cite{MATLAB:2020} to model the behaviour of realisable TESs under a range of particle measurement conditions, including factors such as detector noise, response time, and particle flux. The results were then compared with analytical expressions for key performance indicators. The simulation method proceeds as follows: the dynamical behaviour of the TES is used to calculate the recorded signal when a flux of particles having known energies is incident on the device. The recorded output is then passed to representative analysis software, which calculates electron arrival times and their corresponding energies. The simulations take into account noise in the device, and factors such as pile-up. As the incident particle energies are known exactly, the ability of both the detector and the analysis method to measure spectra under a variety of conditions can be studied. Moreover, using Fisher information, the fundamental energy resolution available in the simulated dataset can be determined, irrespective of the data analysis method used, to decouple the contributions from the detector and the specific analysis method chosen. In this paper we describe the simulations in detail, apply the method to the specific case of X-ray photoelectron spectroscopy, and evaluate the capabilities of TESs as XPS spectrometers by comparing their energy resolutions and measurement rates with traditional XPS analysers.

\section{\label{sec:level2}TES Calorimetry Theory}

Transition-edge sensors comprise a particle absorber thermally connected to a superconducting thin film cooled to within its superconducting transition region. In some TESs, the superconducting film itself can act as the absorber. The addition of heat into the superconducting film from a particle absorption event results in a measurable electrical response in the TES by way of an electrothermal feedback loop. The electrothermal feedback loop is established by voltage-biasing the TES such that Joule heating in the superconductor maintains the TES at a stable operating temperature within the superconducting transition region \cite{Irwin2005}. The resistance of the superconductor is highly temperature-dependent so the addition of external thermal energy into the TES increases the resistance, reducing current flow and Joule heating, maintaining the TES at its equilibrium temperature. Changes in input power can be monitored by measuring the TES current flow. 

The thermal behaviour of a TES is described by
\begin{equation} \label{eq1}
C\frac{dT(t)}{dt}=P_v(t)-P_b(t)+P_{abs}(t),
\end{equation}
where $C$ is the overall heat capacity, $P_v$ is the Joule heating power, $P_{abs}(t)$ characterises the thermal power from particle absorption and $T(t)$ is the time-dependent temperature. Assuming perfect voltage biasing, the Joule heating is given by $P_v = V_0^2/R(T)$ where $V_0$ is the bias voltage and $R(T)$ is the TES resistance. The heat flow between the TES and bath can be modelled as $P_b = K\left(T^n-{T_b}^n\right)$ where $K$ is a constant, $T_b$ is the bath temperature, and $n$ is a parameter determined by the dimensionality of thermal conductivity between the device and bath, typically between 2 and 4. Electrothermal feedback maintains the TES temperature very close to the superconducting transition, $T_c$. Assuming instantaneous particle thermalisation, solving Eqn.\,\ref{eq1} in the small signal limit, such that $P_{abs}<<P_v$, gives the current response to be (see Appendix) \cite{figueroa2001theory}
\begin{equation} \label{eq2}
\Delta I(t) = -\frac{E_{abs}}{C}\frac{\alpha I_0}{T_c}e^{-t/\tau_{\text{eff}}},
\end{equation}
where $\Delta I(t)$ is the difference in current passing through the TES from its equilibrium current, $I_0$, $E_{abs}$ is the absorbed particle energy, and $\tau_{\text{eff}}$ is the effective TES response time (see Appendix). The factor $\alpha$ is a measure of the sharpness of the superconducting transition and is given by
\begin{equation} \label{eq3}
\left.\alpha \equiv \frac{T_c}{R} \frac{\partial{R(T)}} {\partial{T}}\right|_{I_0}.
\end{equation}

The accuracy and precision with which particle energies can be determined depends on noise in the measurement system. For a well-designed TES, the main noise sources are phonon noise in the coupling to the heat bath and Johnson noise in the biased superconducting film. The impact of these noise sources can be characterised using noise-equivalent power, NEP.  A detector's NEP is the amount of power that must be applied to the device to achieve a signal-to-noise ratio of unity in one hertz readout bandwidth. The contributions of phonon and Johnson noise to the NEP are \cite{Irwin2005}
\begin{equation} \label{eq4}
    \text{NEP}_{ph}=\sqrt{4k_BT_c^2G F\left(T_c,T_b\right)}
\end{equation}
and
\begin{equation} \label{eq5}
    \text{NEP}_{J}(\omega)=\sqrt{4k_BT_cP_v\left(\frac{GT_c}{\alpha P_v}\right)^2(1+\omega^2\tau^2)},
\end{equation}
where $k_B$ is the Boltzmann constant, $G$ is the thermal conductance between the device and the bath, given by $G = nKT_c^{n-1}$, $\omega$ is the angular frequency, and $\tau=C/G$ is the TES response time in the absence of electrothermal feedback. The function $F(T_c,T_b)$ characterises the nature of thermal power flow between the TES and bath, and can be parameterised such that $F(T_c,T_b) = \gamma$, where $\gamma$ typically has a value between 0.5 and 1. Adding these noise sources in quadrature and simplifying, assuming zero bath temperature, gives
\begin{equation} \label{eq6}
    \text{NEP}(\omega)=\sqrt{4k_BT_c^2G\left(\frac{n}{\alpha^2}(1+\omega^2\tau^2)+\gamma\right)},
\end{equation}
representing the combined NEP due to Johnson and phonon noise. This result is further simplified under the condition that $\omega <<1/\tau$, whereupon the frequency dependence can be neglected and the noise becomes white.

The energy range that can be measured is determined by the TES's heat capacity. If the incident energy is too great, the temperature will rise outside the superconducting transition window and the TES will enter the normal metal state. The corresponding saturation energy, $E_{sat}$, is approximated as \cite{figueroa2001theory}
\begin{equation}\label{eq7}
        E_{sat}\approx\frac{CT_c}{\alpha},
\end{equation}
with $T_c$ being the superconductor's critical temperature.
If the TES's temperature increases towards the edge of the superconducting transition window, the behaviour becomes non-linear and more sophisticated analysis is required to determine particle energies.

\section{\label{sec:level3}Simulation Model and Analysis}

\begin{figure}
\centering
  \begin{tabular}{@{}p{0.9\linewidth}}
    \subfigimg[width=\linewidth]{a)}{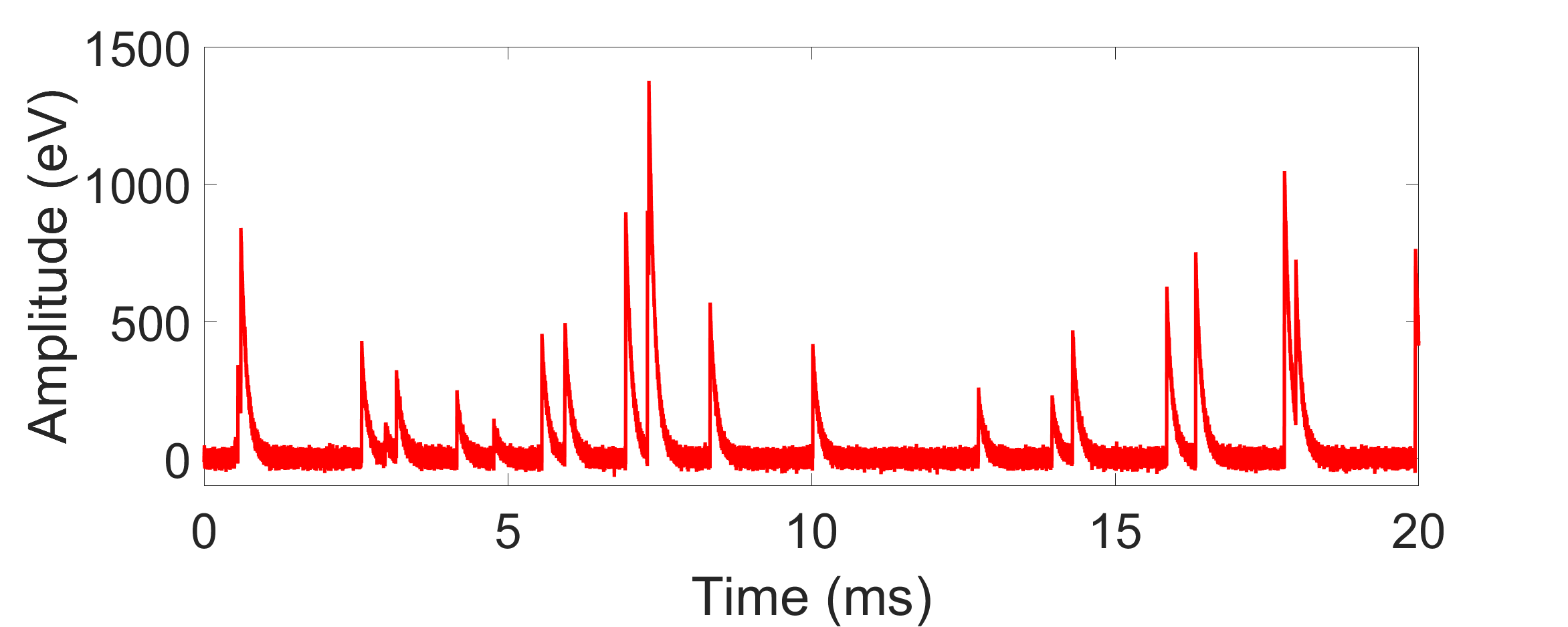}
    \subfigimg[width=\linewidth]{b)}{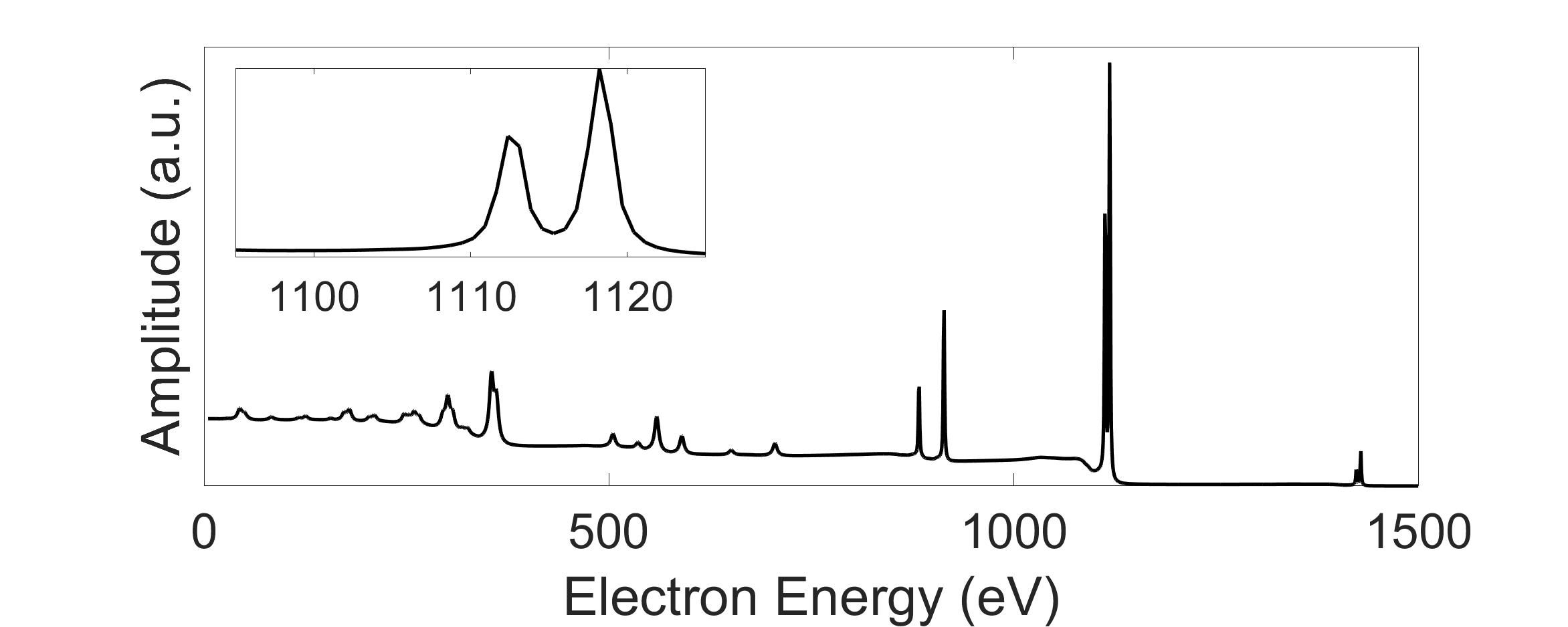}
    \subfigimg[width=\linewidth]{c)}{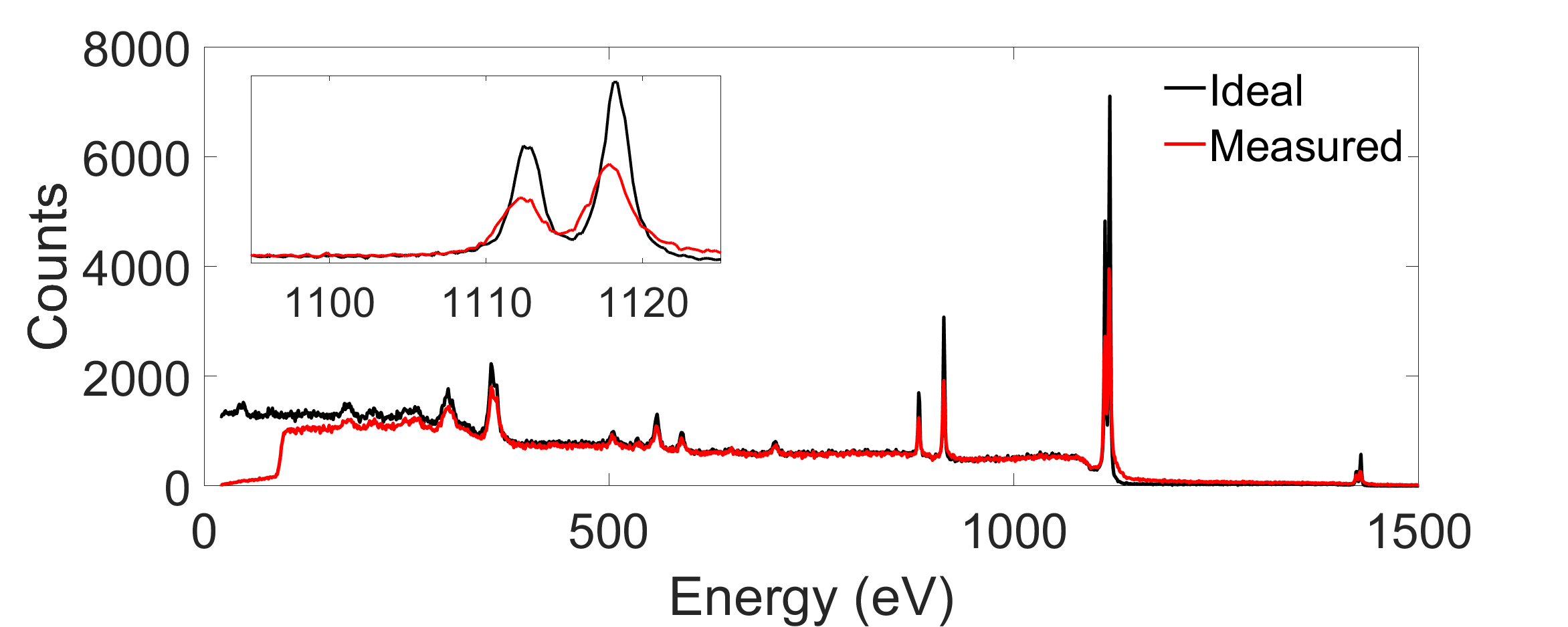}
\end{tabular}
\caption{a) Simulated response of an TES calorimeter to a stream of incident electrons with energies sampled from (b). b)  XPS spectrum of silver calculated using SESSA software \cite{smekal2005simulation}. c) Reconstructed XPS spectra using the simulated calorimeter data. The black line shows the behaviour in an ideal case, where the energy of each electron is determined perfectly. The measured line in red shows what was achieved by identifying events using the described analysis method. A minimum event threshold of 100\,eV was implemented in the analysis resulting in the sharp reduction of counts below 100\,eV.}
\label{fig:Sim}
\end{figure}

The first step in the simulation software was generating a set of particle arrival times and corresponding energies. The arrival times were either randomly sampled from a uniform probability distribution with a given mean particle flux, or chosen with a fixed interval between arrivals to examine behaviour free from overlapping particle absorption events. Particle energies were randomly sampled from a specified input energy spectrum using inverse transform sampling. A discrete time series dataset was then built with a given sampling period over a predetermined measurement duration. Given the assumptions that the TES is linear, calibrated, and not saturated by the signal, it is not necessary to define the values of $C$, $\alpha$, $I$ and $T$ in Eqn.\,\ref{eq2}, allowing the signal amplitude to refer to energy directly. Particle absorption events were placed in the time series as single data points at the discretised time immediately following the particle arrival time with the amplitudes of these data points being the particle energies. The data set was converted to the TES response by convolution with Eqn.\,\ref{eq2} with unit amplitude such that
\begin{equation}\label{eq8}
        y_i= \sum_{j=-\infty}^{\infty} x_{j} e^{t_{i-j}/\tau_{\text{eff}}},
\end{equation}
where $y_i$ is the TES response and $x_i$ is the input data. As Eqn.\,\ref{eq2} is linear, the sign has no practical bearing on the analysis so the TES response sign has been taken as positive rather than negative for simplicity of representation. White noise was generated by randomly sampling values from a Gaussian distribution, scaling by a given noise amplitude and then adding the result to the TES response. Fig.\,\ref{fig:Sim}a shows a segment of generated data with electron energies being sampled from the XPS spectrum of silver (shown in Fig.\,\ref{fig:Sim}b). Silver is a common XPS reference material and so has been used to characterise the simulated TES performance.

Following device simulation, the data were then analysed and used to reconstruct the spectrum of the electron spectrum flux. The data analysis method can be divided into two parts: arrival time estimation and pulse amplitude estimation. The entire dataset was first divided into segments, with each segment containing either one particle event or multiple overlapping events. The segments were created using two threshold values: a minimum event threshold and a noise threshold. If the signal rises above a chosen minimum event threshold, a particle absorption event was said to have occurred and a data segment was made. The start and end times of the segment were determined by the nearest times before and after the minimum event threshold crossing where the signal crosses a chosen noise threshold. Following this procedure, every data segment would begin with the signal amplitude rising above the noise threshold, crossing the minimum event threshold, and ending after falling below the noise threshold. Therefore, each segment can contain multiple particle absorptions if the TES responses overlap in time. The minimum event threshold is set by the minimum particle energy of interest (or sufficiently above the noise floor to reduce false triggers to acceptable levels). The noise threshold should lie within the noise floor to minimise truncation of the detector response, both on the leading and falling edges.

With the data parsed into smaller segments, each segment was analysed to find the number of electron arrivals within the segment and their respective arrival times. To improve event discrimination and timing, the data was passed through a matched filter. The matched filter convolves the data with a time-reversed form of the desired signal: the signal being Eqn.\,\ref{eq2} in this case. The matched filter maximises signal-to-noise with particle absorption events becoming distinct peaks. After applying the filter, the number of peaks within each segment was found and particle arrival times were recorded as the time of the peak maximum.

Knowing the number of particles and their arrival times, the corresponding particle energies were calculated. This was achieved using maximum likelihood estimation, which in the presence of Gaussian noise reduces to least squares estimation. The ideal model describing the measured data is a superposition of exponential decays with different arrival times and amplitudes. For a dataset with $n$ sampled points and $N_p$ particle events, the i'th datapoint is given by
\begin{equation}\label{eq9}
    \mu_i = \sum_{j=1}^{N_p} \frac{E_j}{\tau_{\text{eff}}} \exp{
    \left(\frac{-(t_i-\phi_j)}{\tau_{\text{eff}}}\right)
    }H(t_i,\phi_j),
\end{equation}
where $H(t_i,\phi_j)$ is a step function with a value of 0 when $t_i<\phi_j$, and 1 when $t_i>\phi_j$; $\phi_j$ is the arrival time of the $j$'th electron and  $E_j$ is the corresponding particle energy. Consider a vector of signal amplitudes, $\textbf{y}$, containing $n$ sampled datapoints denoted $y_i$. If $\textbf{y}$ possesses Gaussian noise, then the probability of observing $\textbf{y}$  according to Eqn.\,\ref{eq9} is
\begin{eqnarray}\label{eq10}
    P(\textbf{y}|\{E_j,&&\phi_j\},\sigma) = \nonumber\\
    &&\left(2\pi\sigma^2\right)^{-N/2} \exp\left(\sum_{i=1}^n
    \frac{-(y_i-\mu_i)^2}{2\sigma^2}  \right).
\end{eqnarray}
The resulting log-likelihood function becomes
\begin{eqnarray}\label{eq11}
    l(\{E_j\}|\textbf{y},\{\phi_j\},\sigma) = &&N\text{log}
    \left(\frac{1}{\sqrt{2\pi\sigma^2}}\right)\nonumber\\
    && + \frac{1}{2\sigma^2}\sum_{i=1}^n-(y_i-\mu_i)^2.
\end{eqnarray}
The energy estimates, $\Tilde{E_j}$, are the values that maximise the log-likelihood function such that
\begin{equation}\label{eq12}
\dfrac{\partial l}{\partial E_j}=0.
\end{equation}
By applying this condition to Eqn.\,\ref{eq11}, it follows that
\begin{equation}\label{eq13}
    y_i
    = \sum_{j=1}^{N_p} \frac{\Tilde{E_j}}{\tau_{\text{eff}}} x_{ij},
\end{equation}
where $x_{ij} = \exp{\left(\frac{-(t_i-\phi_j)}{\tau_{\text{eff}}}\right)}H(t_i,\phi_j)$. Eqn.\,\ref{eq13} can be rewritten in matrix form,
\begin{equation}\label{eq14}
    \textbf{y} = \frac{1}{\tau_{\text{eff}}} \textbf{Xa},
\end{equation}
where $\textbf{y}$ is the vector of measured data, $\textbf{X}$ is the matrix populated by $x_{ij}$, and $\textbf{a}$ is the vector of particle energies. The set of energies was estimated from Eqn.\,\ref{eq14} by calculating the QR decomposition of $\textbf{X}$. An example spectrum produced by this method is compared with the source spectrum in Fig.\,\ref{fig:Sim}c. The source spectrum was calculated using the energies and arrival times of the electrons in the incident beam, and should therefore be regarded as the desired result from the TES measurements.

The difference between the recovered spectrum, in Fig.\,\ref{fig:Sim}c, and the source spectrum from which the particle energies were sampled, in Fig.\,\ref{fig:Sim}b, can be linked to three different effects: (i) the finite number of measurements, (ii) pileup, and (iii) energy estimation errors, which occur in the presence of noise. The effects of pileup are most clearly visible around 200 eV and 1200 eV. Events where a high-energy particle overlaps with a low-energy particle often results in a failure to identify the low energy response leading to a single particle being measured with excess energy, removing low energy counts and pushing the high energy measurement to even higher energies.

A central question is to what extent the above limitations are inherent in the behaviour of TESs, in particular noise, and to what extent are they artifacts of the specific data analysis method used? The presence of noise in any system will limit the achievable resolution. One method of determining the resolution limit is via Fisher information analysis. Fisher information is a measure of the information contained in a random variable with respect to a set of assumed model parameters: in our case electron energy. Fisher information is defined as
\begin{equation}\label{eq15}
    I(\theta) = -\mathbb{E}\left[
    \left(
    \frac{\partial}{\partial \theta}l(\theta|x)
    \right)^2\middle|\theta
    \right],
\end{equation}
where $l(\theta|x)$ is the log-likelihood function with parameter $\theta$ and variable $x$. A key aspect of Fisher information is the Cramer-Rao bound which limits the measurement precision of $\theta$ to
\begin{equation}\label{eq16}
\text{Var}(\tilde{\theta}) \geq \frac{1}{I(\theta)},
\end{equation}
using $\tilde{\theta}$ to denote the estimated value of the true parameter $\theta$.

Using Eqn.\,\ref{eq10} and Eqn.\,\ref{eq11}, the measurement of a single particle arriving at time zero with energy $E$ has Fisher information of
\begin{equation}\label{eq17}
    I(E) =\frac{1}{\sigma^2 \tau_{\text{eff}}^2}\sum_{i=1}^{n}\exp\left(\frac{-2t_i}{\tau_{\text{eff}}}\right),
\end{equation}
and if the sampled data points are equally spaced in time with period $\Delta t$, Eqn.\,\ref{eq17} can be simplified to
\begin{equation}\label{eq18}
I(E)=\left[\sigma^2 \tau_{\text{eff}}^2\left(1-\exp\left(\frac{-2\Delta t}{\tau_{\text{eff}}}\right)\right)\right]^{-1}
\end{equation}
through the use of a geometric series summation. Rather than using the noise amplitude $\sigma$, it is more useful to characterise noise using NEP; in the case of Gaussian white noise,
\begin{equation}\label{eq19}
\sigma = \frac{\text{NEP}}{\sqrt{2\Delta t}}.
\end{equation}
Substituting Eqn.\,\ref{eq18} and Eqn.\,\ref{eq19} into Eqn.\,\ref{eq16} provides the limiting error
\begin{equation}\label{eq20}
\Delta E=2.355\frac{\text{NEP}}{\sqrt{2\Delta t}}\tau_{\text{eff}}\sqrt{1-\exp\left(\frac{-2\Delta t}{\tau_{\text{eff}}}\right)},
\end{equation}
or in the continuous case
\begin{equation}\label{eq21}
\Delta E=2.355\,\text{NEP}\sqrt{\tau_{\text{eff}}},
\end{equation}
where $\Delta E$ denotes the standard deviation of particle energy. The factor 2.355 comes from converting standard deviation to full-width at half-maximum. The energy resolution limit in Eqn.\,\ref{eq21} is in agreement with the result from \cite{3M} when considering white noise.

\begin{figure}
\includegraphics[width=8cm]{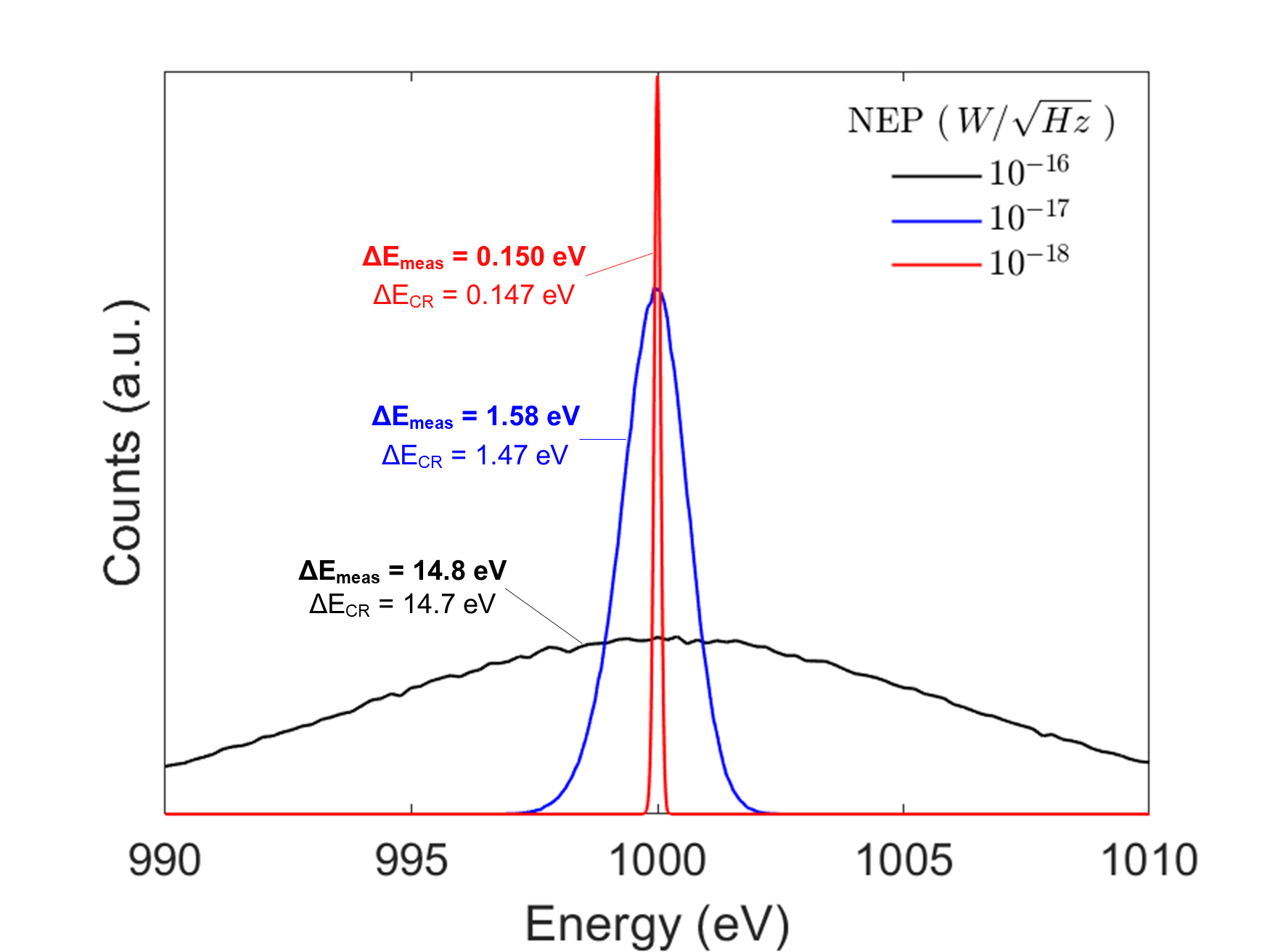}
\centering
\caption{Measured energy spectra obtained by analysing simulated streams of incident electrons with periodic particle arrival times to prevent pileup and with uniform energies. Each simulation was performed with a detector response time of 100 $\mu\,$s and at different noise-equivalent powers (NEP). The resulting measured FWHM, $\Delta E_\text{meas}$ is compared with the Cramer-Rao lower bound $\Delta E_{CR}$ from Eqn.\,\ref{eq21}.}
\label{fig:Fisher}
\end{figure}

Fig.\,\ref{fig:Fisher} compares the predicted energy resolutions from Eqn.\,\ref{eq21} against a set of three simulated measurements. For each simulation, a series of 1000 eV particles were generated with arrivals spaced 10\,$\tau_{\text{eff}}$ apart in time to remove any overlap. The simulation was run for three different NEP's and the resulting energy resolutions ($\Delta E_\text{meas}$) were calculated and compared to the Cramer-Rao lower bounds ($\Delta E_{CR}$). With every NEP simulated, the obtained energy resolutions approach the Cramer-Rao limits of 14.7\,eV, 1.47\,eV and 0.147\,eV for NEPs of $10^{-16}$, $10^{-17}$ and $10^{-18}\,\text{W}/\sqrt{\text{Hz}}$, respectively, demonstrating excellent performance of the event identification and least squares analysis. For this reason we believe that the data analysis method we have chosen gives, over the range of parameters considered, results that are truly reflective of the fundamental behaviour of the device.

\section{\label{sec:level4}TES as X-ray Photoelectron Spectrometers}

The most common instrument for XPS measurements is the hemispherical analyser (HSA). The energy filter consists of a lens assembly that decelerates and focuses electrons onto an entrance slit of width $w$. The electrons are deflected between two concentric, charged hemispherical electrodes to the detector. The potential difference between the hemispheres determines the pass energy ($E_p$). Electrons having this kinetic energy are deflected to the centre of the detector. Electrons with slightly different energies are deflected to the sides of the detector allowing a small range of energies to be detected concurrently. The FWHM energy resolution of a hemispherical analyser is approximately given by \cite{PracticalSurfaceAnalysis}
\begin{equation} \label{a1}
\Delta E = E_p\frac{w}{2R},
\end{equation}
where $R$ is the hemispherical radius. As such, the energy resolution depends on the width of the entrance slit, and pass energy used. Using a hemispherical radius of 200\,mm, a slit width of 0.1 to 5 mm and a pass energy of 200\,eV  (typical of current XPS analysers), the energy resolution ranges from 0.1\,eV to 5\,eV.

TESs have no such relationship between the energy resolution and count rate; instead, the optimal count rate will occur at the maximum electron flux before the effects of particle pileup degrade measurement quality beyond the threshold needed for a particular application. This pileup limitation can be improved by developing lower response time detectors to reduce the incidence of pileup, or by developing arrays of TESs for concurrent measurement.

TES energy resolution is obtained by substituting Eqn.\,\ref{eq6}, the expression for NEP, into Eqn.\,\ref{eq21} to provide
\begin{equation}\label{eq23}
    \Delta E \geq \sqrt{4k_BT_c^2G\tau_{\text{eff}}\left(\frac{n}{\alpha^2}+\gamma\right)},
\end{equation}
in the case of white noise. In TESs with strong electrothermal feedback, such that the majority of thermal energy is removed via electrothermal feedback rather than flowing to the thermal bath, the detector time constant is $\tau_{\text{eff}} = {nC}/{\alpha G}$. With $n<<\alpha$ and $\gamma = 0.5$, Eqn.\,\ref{eq23} can be rewritten as
\begin{equation}\label{eq24}
\begin{split}
    \Delta E &\geq 2.355\sqrt{4k_BT_c^2\frac{C}{\alpha}\left(\frac{n}{2}\right)}\\
             &\geq 2.355\sqrt{4k_BT_cE_{sat}\left(\frac{n}{2}\right)}.
\end{split}
\end{equation}

The minimum saturation energy needed in an XPS application is determined by the X-ray source used; Al K$\alpha$ is commonly used with an energy of 1486.6\,eV. Eqn.\,\ref{eq24} shows that the best energy resolution possible for a TES at 200\,mK device temperature, with $n$=2, $\alpha$=100, and a saturation energy of 1500\,eV, is approximately 0.8\,eV FWHM. Detectors targeting these higher energy ranges and resolutions are no longer described by the small-signal limit and as such would require more sophisticated data analysis methods than the one described here. Nevertheless, this can be done through pulse-template matching techniques, which are used with X-ray TES calorimeters.  Alternatively, the TES can be redesigned to give a higher saturation energy, compared with the parameters used here, but this would degrade the noise and lead to low energy resolution. In fact, the combination of energy range and resolution correspond strongly with the requirements of X-ray calorimetry where TESs have already found a number of applications  \cite{Ullom2015review,Athena2016,Lee2019Xray}.

The TES energy resolution of 0.8\,eV lies within the range of 0.1\,eV to 5\,eV achieved with hemispherical analysers (HSA) at 200\,eV pass energy. However, to achieve measurements with 0.1\,eV resolution and a 1500\,eV saturation energy would require a TES temperature of approximately 5\,mK which is too low for conventional XPS instruments.

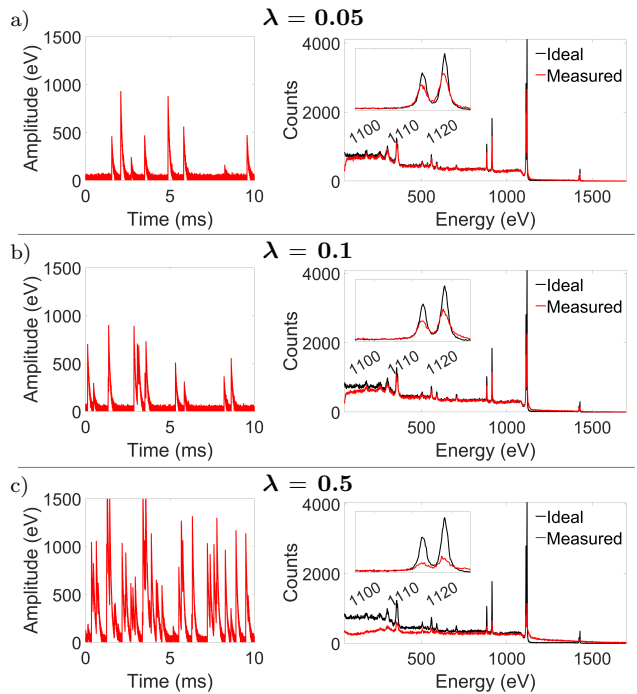
\begin{figure}
\centering
\begin{tikzpicture}
\node[anchor=south west,inner sep=0] (image) at (0,0) {
  \begin{tabular}{@{}p{0.4\linewidth}@{}p{0.59\linewidth}}
    \subfigimg[width=\linewidth]{}{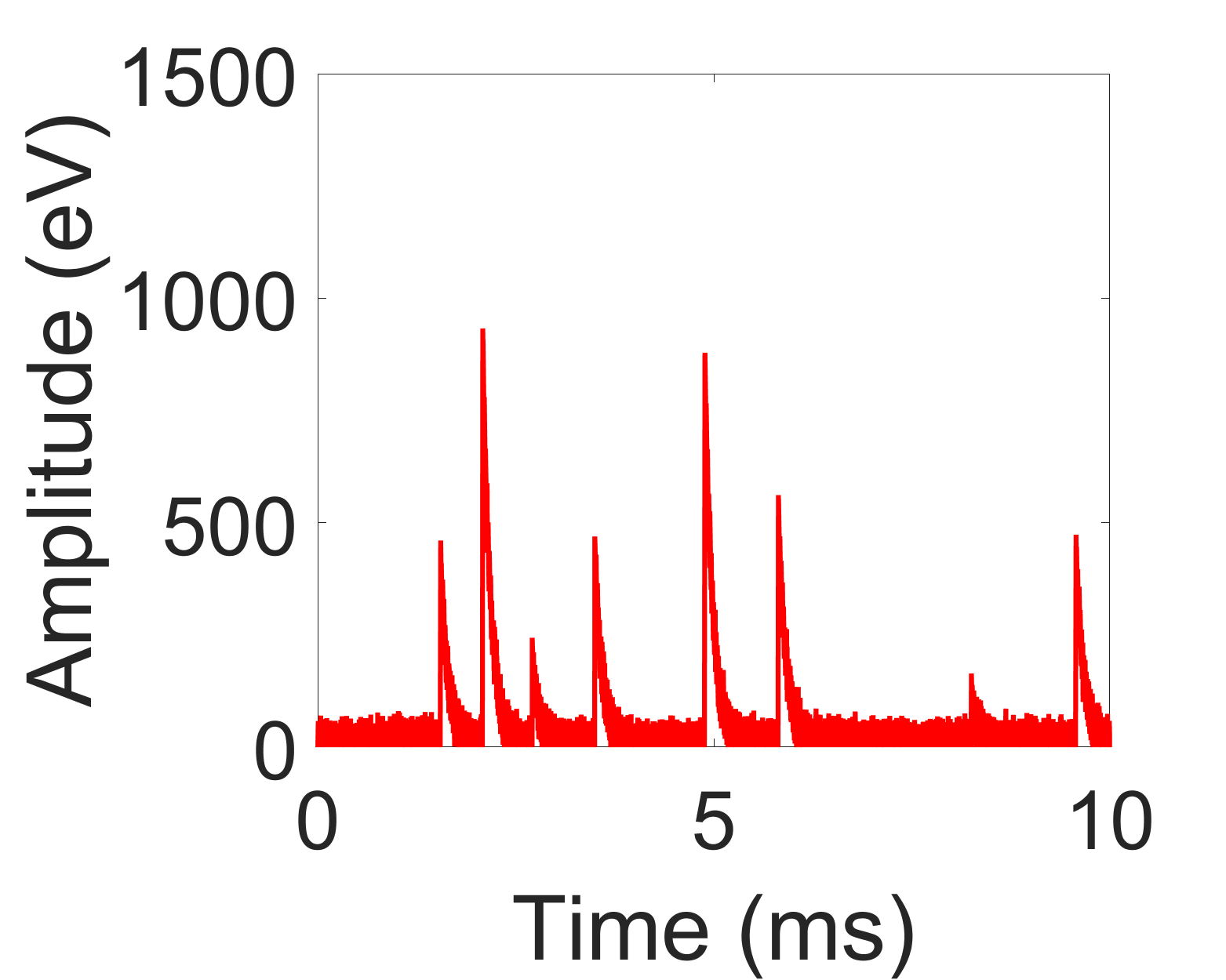} &
    \subfigimg[width=\linewidth]{}{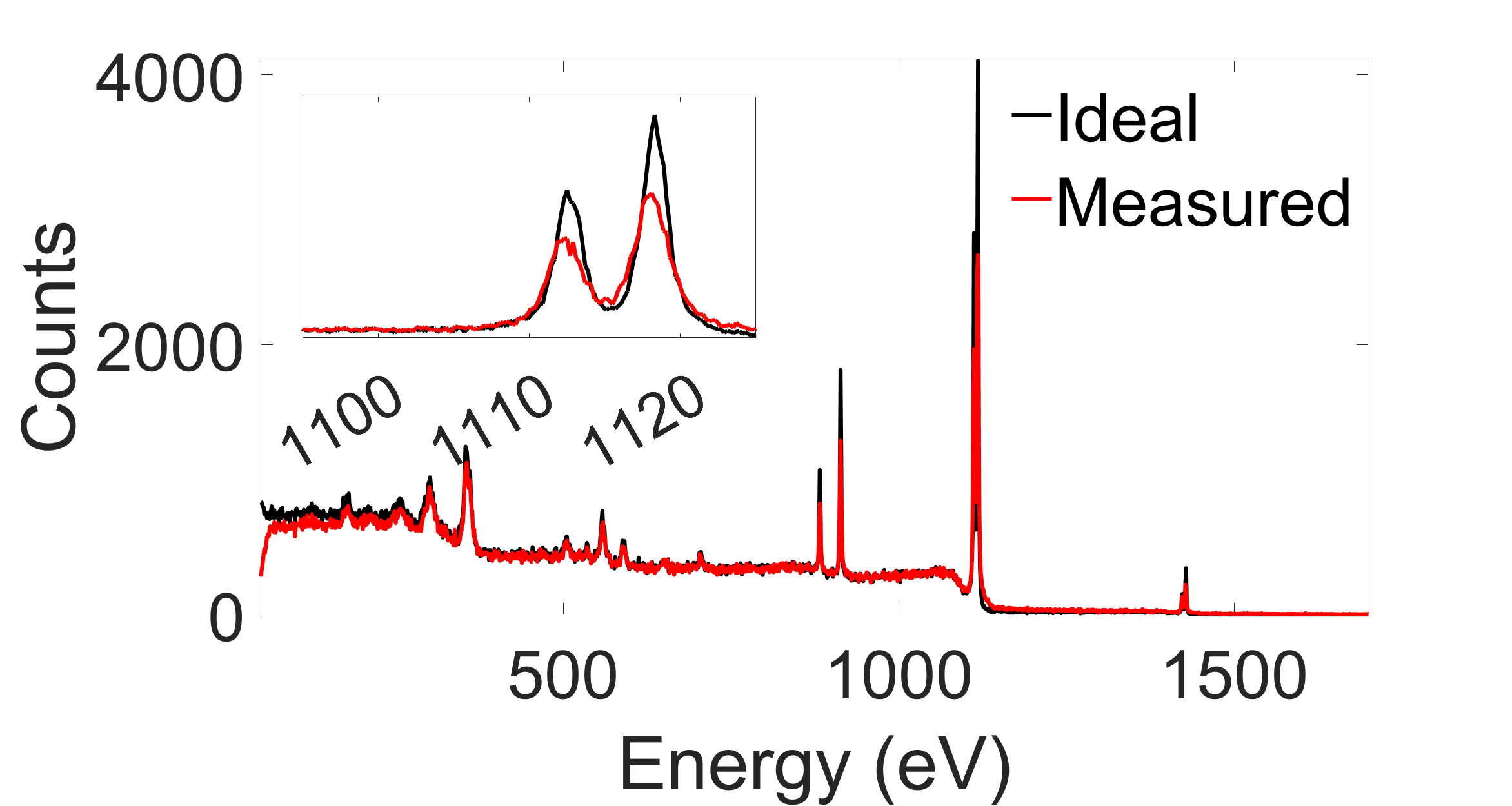}\\[2mm]
    \subfigimg[width=\linewidth]{}{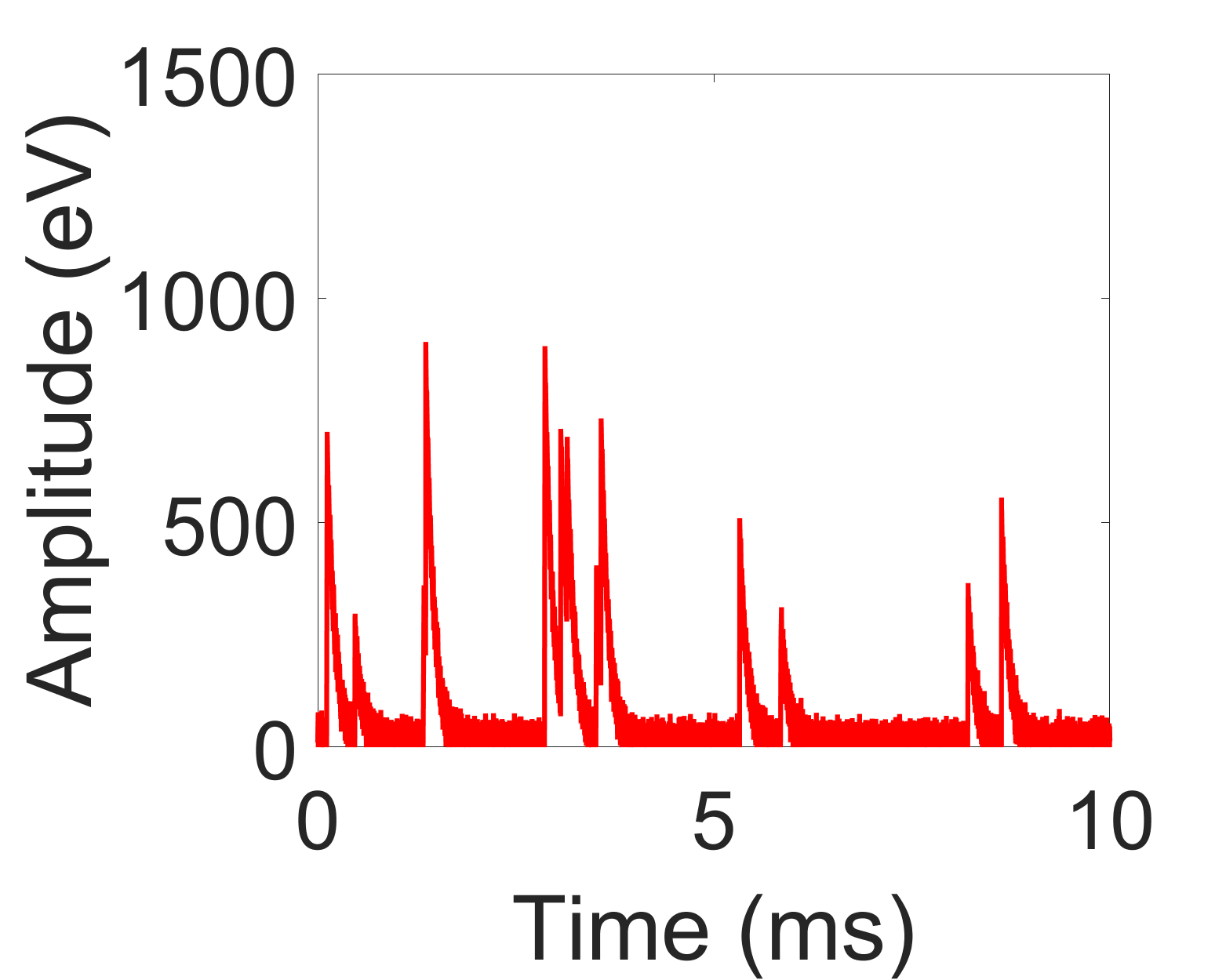} &
    \subfigimg[width=\linewidth]{}{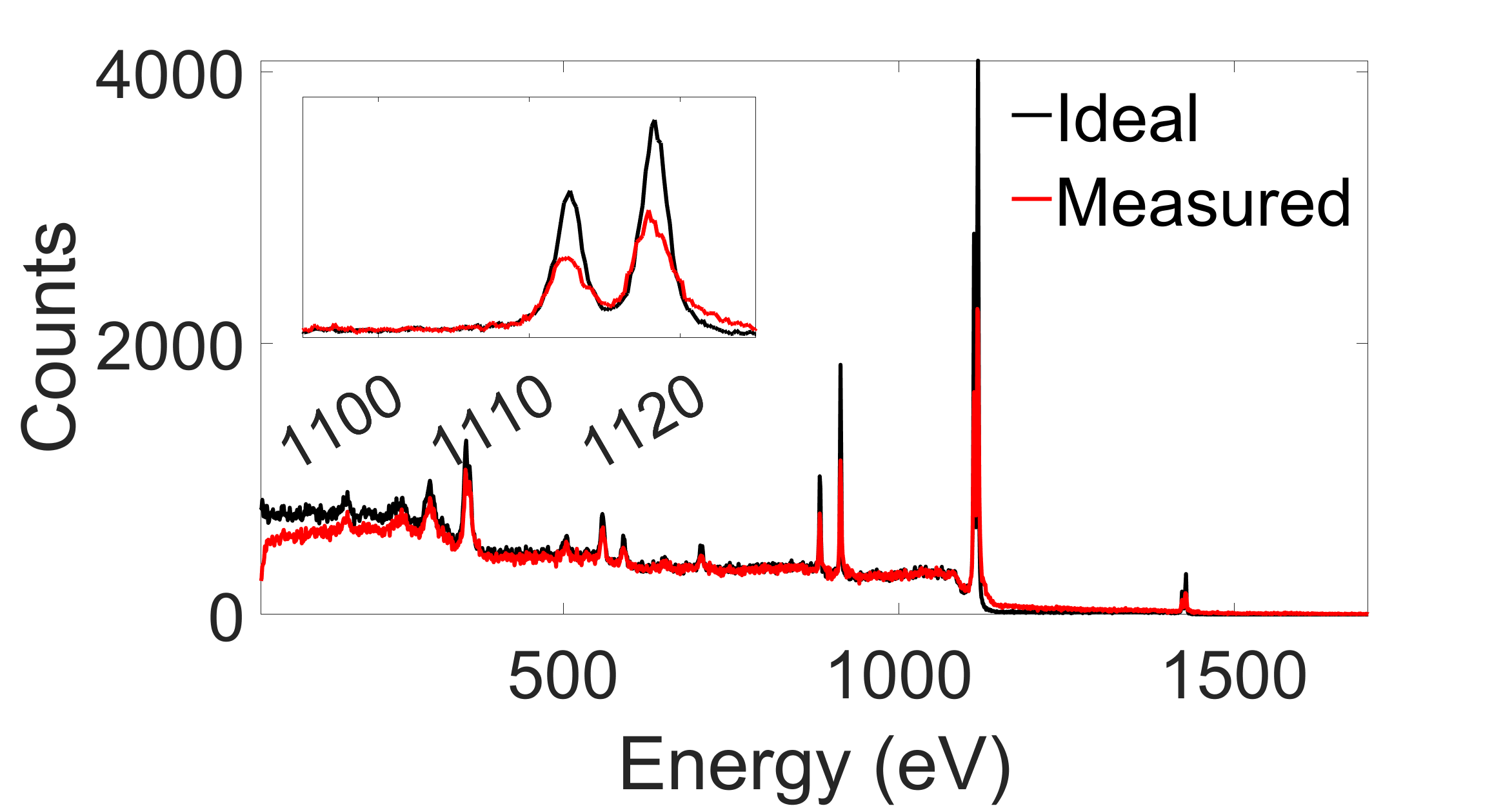}\\[2mm]
    \subfigimg[width=\linewidth]{}{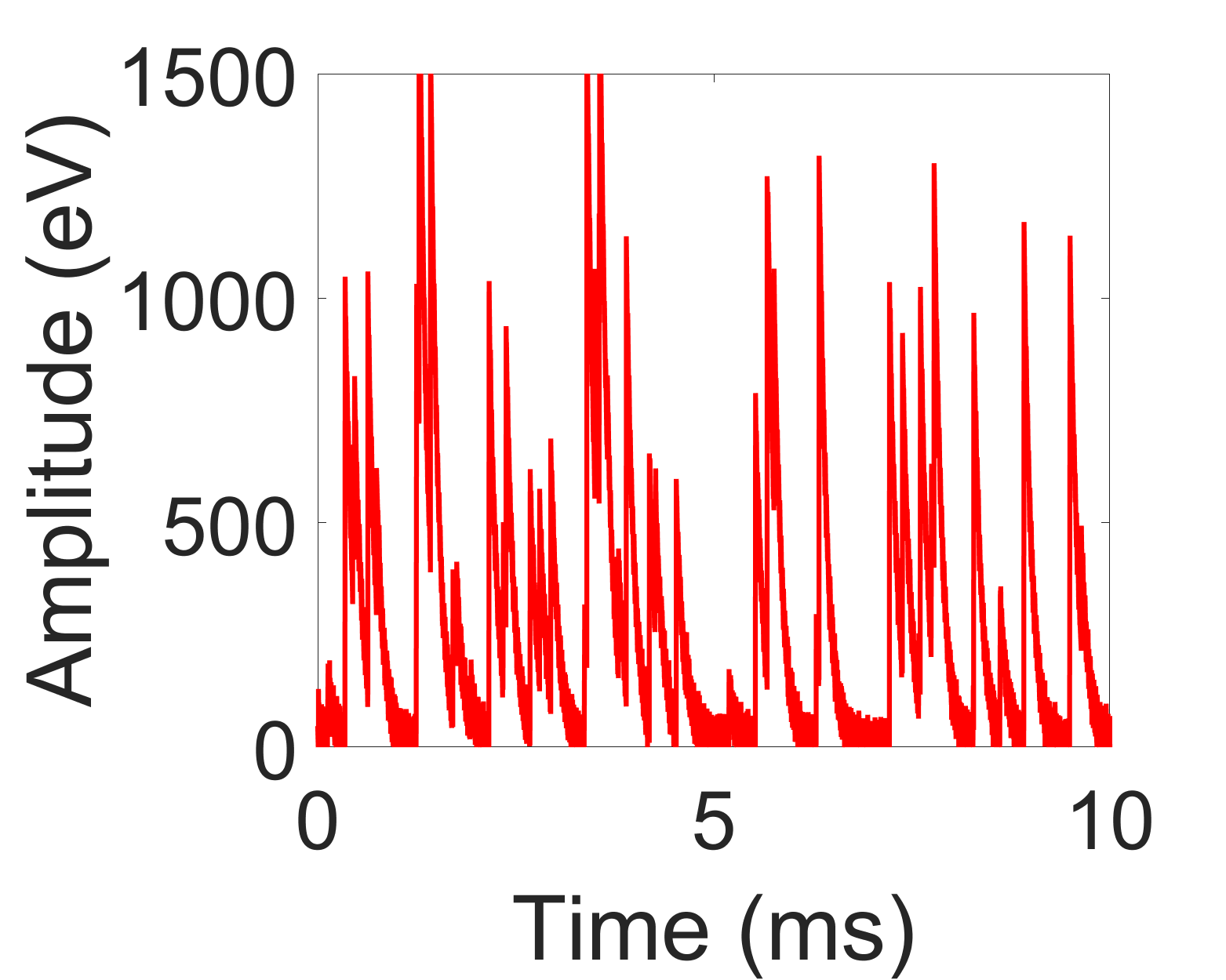} &
    \subfigimg[width=\linewidth]{}{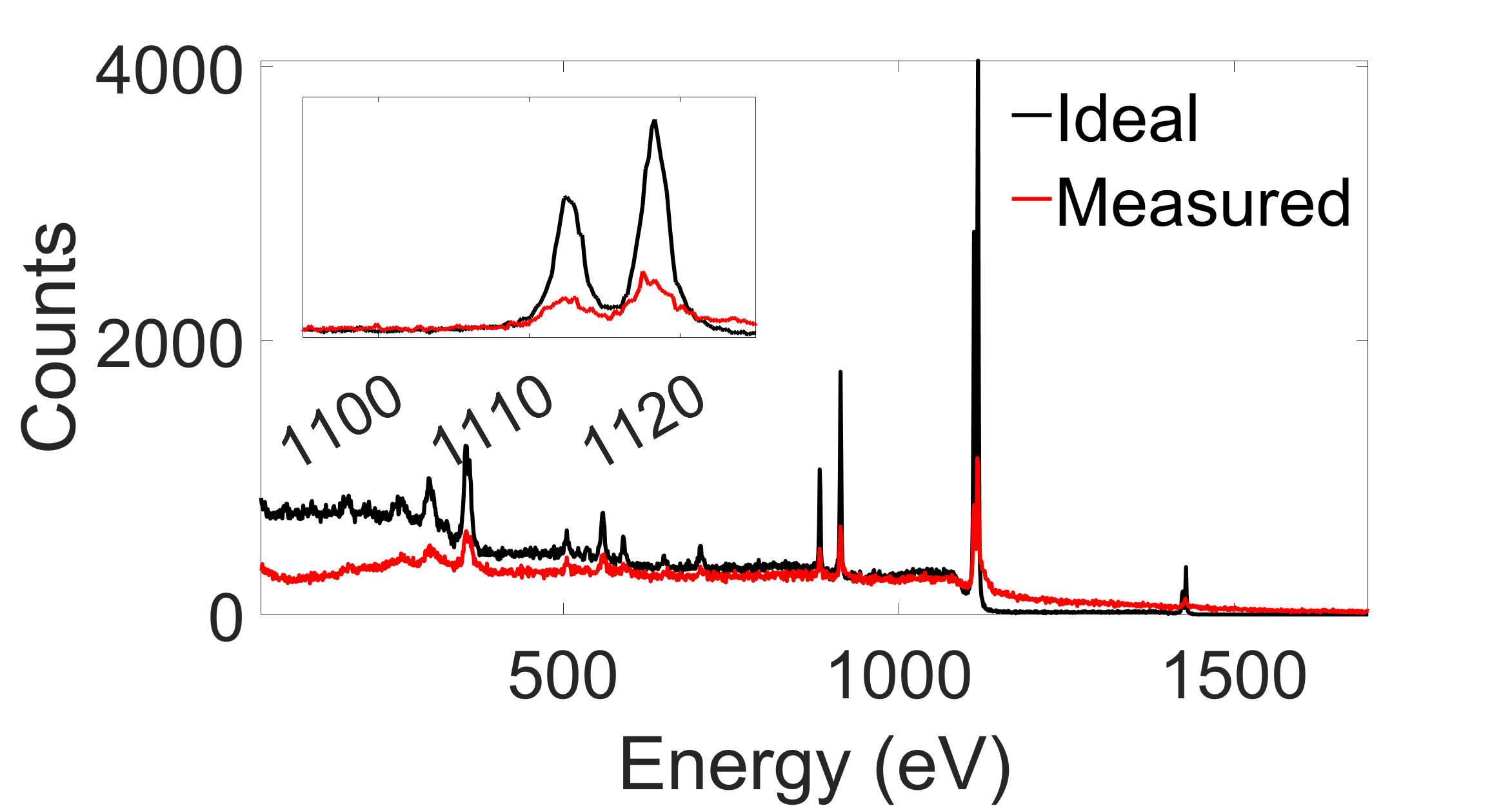}
\end{tabular}
};
\draw [gray, thin] (0,3.1) -- (8.2,3.1);
\draw [gray, thin] (0,6.2) -- (8.2,6.2);
\node[text width=1.5cm] at (4,9.1) {$\boldsymbol{\lambda}=\textbf{0.05}$};
\node[text width=1.5cm] at (4,6.) {\textbf{$\boldsymbol{\lambda}=\textbf{0.1}$}};
\node[text width=1.5cm] at (4,2.9) {\textbf{$\boldsymbol{\lambda}=\textbf{0.5}$}};

\node[text width=.3cm] at (0.05,9.05) {\footnotesize a)};
\node[text width=.3cm] at (0.05,5.95) {\footnotesize b)};
\node[text width=.3cm] at (0.05,2.85) {\footnotesize c)};
\end{tikzpicture}

\caption{Demonstration of the effect of particle pileup on the achieved spectral resolution. The value $\lambda$ is the effective TES response time multiplied by the particle arrival rate and represents the relative incidence of pileup events. a) At low values of $\lambda$, pileup is infrequent and the true spectrum can be reproduced accurately b) $\lambda$ = 0.1 provides a compromise between high count rate and acceptable levels of spectral degradation due to pileup. High levels of pileup as shown in c) severely degrades the measured spectrum, especially at low energies where these events are often obscured by higher energy particles. Each simulation has been performed with the same number of incident electrons.}
\label{fig:Pileup}
\end{figure}

The last property to consider is measurement rate, determined by the TES's response time. Using the simulation in Sect.\,3, the electron flux that best balances high count rate and pileup can be empirically determined. Fig.\,\ref{fig:Pileup} compares three simulations, each with the same number of incident electrons, but with different electron fluxes. The relative occurrences of event pileup is denoted using the factor $\lambda$, which is equal to the TES response time, $\tau_{\text{eff}}$ multiplied by the electron arrival rate. Increasing either the response time or the electron flux by a given factor increases the rate of pileup proportionately. At low pileup, when $\lambda=0.05$, the measured spectrum is accurately reproduced, but the detector is idle for extended periods of time, unnecessarily raising the measurement time needed to produce such a spectrum. At high pileup, ($\lambda=0.5$), the measured spectrum is heavily distorted. Fig.\,\ref{fig:Pileup}b provides the balance between high count rates and measurement accuracy, with twice the measurement speed as compared with Fig.\,\ref{fig:Pileup}a, and the most prominent errors occurring at very low energies. For XPS spectra, low energy data below 200\,eV are of little interest and are not typically measured whereas the characteristic peak energies and relative amplitudes are highly important. The observed pileup at $\lambda=0.1$ has little detrimental impact on the measured spectrum unless quantitative compositional data is needed at which point either a lower electron flux should be used or pileup rejection is implemented where overlapping events are identified and ignored.
As the electrons have random arrival times, the proportion of overlapping events can be calculated using Poisson statistics. The probability of $n$ events occurring in an interval of duration $t$ is
\begin{equation}\label{eq25}
    P(n) = \frac{\left(kt\right)^ne^{-kt}}{n!}
\end{equation}
where $k$ is the average particle flux. Given a particle absorption event has occurred, the probability that no particles are absorbed within time $t$ before and after is therefore $e^{-2kt}$. If non-overlapping are defined as events separated by at least $3\,\tau_{\text{eff}}$, the proportion of events remaining after pileup rejection would be $e^{-6k\tau_{\text{eff}}}$ which is 0.55 when $\lambda = 0.1$ corresponding to a greater count rate than the total count rate at $\lambda = 0.05$. In practice, the efficacy of pileup rejection at improving the quality of the data will be limited by occurrence of multiple particles arriving too closely in time to be distinguished as separate events. Analysing the dataset for Fig.\,\ref{fig:Pileup}b showed that 11\,\% of incident electrons were unidentified due to their proximity to another event with all unidentified pileup events occurring at time differences of 2\,$\tau_{\text{eff}}$ or less.  

A full XPS spectrum measurement with little shot noise would require on the order of $10^{6}$ counts; for reference, the spectrum in Fig.\,\ref{fig:Sim}c was built from $10^{6}$ electrons with a bin size of 1\,eV. A typical HSA would require several minutes to perform such a measurement. For a single TES to measure $10^{6}$ electrons in 5 minutes with a pileup factor of 0.1, the TES response time would have to be 30\,$\mu s$. This response time is faster than response times demonstrated by current TES X-ray calorimeters operating in similar energy ranges, which typically lie around the hundred of microseconds. \cite{Lee2015Fine,Lee2019Xray,Ullom2005}. However, using arrays of TESs, the measurement rate of HSAs can be matched and improved upon. An array of 10 TESs with 120 $\mu$s response times would only require 2 minutes to measure $10^6$ particles with a pileup factor of 0.1. Table \ref{tab:table1} combines the results from this section to provide an example set of TES device parameters for electron spectroscopy and the predicted performance capability of a set of such devices. 

The measurement rate of TES electron spectrometers could be greatly improved by implementing larger arrays. Large arrays of TESs are used in astronomy with sizes ranging up to thousands of devices. TES array size is not limited by fabrication capabilities but instead is limited by the readout system which requires increasingly high-bandwidth, robust multiplexing systems to accommodate the increasing number of devices. In contrast, the readout of small array sizes, on the order of tens of devices as discussed here, is possible using parallelised readout electronics.
\begin{table}
    \caption{\label{tab:table1}%
    Predicted TES performance with the given example set of device parameters where all symbols are as defined in the text. Collection time is the time needed to measure 10$^6$ particles at the given count rate and array size.
    }
\begin{ruledtabular}
\begin{tabular}{ccccccc}&
        \textrm{C}&
        \textrm{$\text{G}_0$}&
        \multirow{2}{*}{\textrm{n}}&
        \multirow{2}{*}{$\mathrm{\alpha}$}&
        $\mathrm{T_c}$&
        \textrm{Array}\\

        & \textrm{\footnotesize (pJ/K)}&
        \textrm{\footnotesize (pW/K)}&
        \textrm{}&
        &
        \textrm{\footnotesize(K)}&
        \textrm{Size}\\
    \colrule \\[-2mm]
        & 0.20 & 50 & 2 & 100 & 0.20 & 10 \\[2mm]

        &
        $\mathrm{\tau_{\text{eff}}}$&
        $\mathrm{E_{sat}}$&
        $\mathrm{\Delta E}$&
         \multirow{2}{*}{$\mathrm{\lambda}$}&
        \textrm{\small Count}&
        \textrm{\small Coll. Time}\\

        & $\mathrm{(\mu s)}$&
        $\mathrm{(eV)}$&
        $\mathrm{(eV)}$&
        $\mathrm{}$&
        \textrm{Rate\,(s$^{-1}$)}&
        \textrm{(min)}\\
    \colrule \\[-2mm]
       & 120 & 2500 & 1.0 & 0.1 & 830 & 2.0\\
\end{tabular}
\end{ruledtabular}
\end{table}

\section{\label{sec:level5}Conclusion}

A simulation platform has been built to investigate the suitability of TESs for electron spectroscopy applications. Fisher information analysis has been applied to the simulated data to determine the fundamental energy resolution limit in measuring particles energies. Fisher information can be extended to more complex scenarios, such as coloured noise, nonlinear energy response or even to quantitatively show the loss of information due to pileup, making it a valuable tool for providing an upper bound limit to detector performance from a given set of experimental conditions.

The simulation platform has been used to examine the potential role of TESs as XPS detectors. The existing design of electron spectrometers uses an energy filtering procedure to perform energy-resolved measurements, which sets up a fundamental trade-off between the instrument’s measurement rate and energy resolution. For high measurement rate applications, such as wide energy range or time-resolved measurements, XPS analysers display energy resolutions worse than 1\,eV which can be improved upon using TESs at 200\,mK.  However, for high-resolution, low count rate measurements, XPS analysers can achieve better than 0.1\,eV resolution which cannot be matched by TESs without using extremely low operating temperatures or greatly reducing the saturation energy. Such devices are however interesting for certain fundamental physics experiments.

The major advantage of using TESs for electron spectroscopy over existing electron spectrometers is the scope for improved measurement rates. Electron spectrometers discard over 99\% of electrons collected by the analyser, which would be measurable using TESs. While an individual TES does not currently have the response times to match the count rates of existing XPS analysers, an array of 10 devices can achieve comparable measurement rates to full-spectrum XPS measurements. By increasing the array size further and using faster detectors, TESs spectrometers can reach orders of magnitude greater measurement times than what is achievable using existing electron spectrometers. Arrays of thousands of TESs are being developed for astronomy, meaning that there is considerable scope for innovation.

\begin{acknowledgments}
This work was supported by EPSRC Cambridge NanoDTC (EP/L015978/1) and the National Measurement System of the UK Department of Business, Energy and Industrial Strategy (project 124089: Metrology for Advanced Coatings and Formulated Products).
\end{acknowledgments}

\appendix
\section{TES Response}
The steps leading from Eqn.\,\ref{eq1} to Eqn.\,\ref{eq2} will be briefly outlined here for the reader's convenience. This outline is drawn from \cite{figueroa2001theory} and \cite{Irwin2005}, and the reader is directed to these references for a more thorough treatment. Solving Eqn.\,\ref{eq1} requires the nature of the Joule heating, heat flow out of the device and the input particle energy to be defined. The absorption of the particle's energy can be assumed to be instantaneous if the internal thermal conductances of the device are very much greater than the thermal conductance from the TES to the bath. In this case, the absorption can be approximated as
\begin{equation} \label{22}
P_{abs}(t) = E_0\,\delta\left(t-t_0\right),
\end{equation}
where the instantaneous nature of the absorption is described by a delta function, $t_0$ is the particle arrival time and $E_0$ is the particle's energy. Heat flow from the TES to the bath can be parameterised as
\begin{equation} \label{a2}
P_b(T) = K\left(T(t)^n-{T_b}^n\right),
\end{equation}
where $K$ is a constant, $T(t)$ is the TES temperature and $T_b$ is the bath temperature. Joule heating is simply described as
\begin{equation} \label{a3}
P_v = \frac{V_0^2}{R(T)}.
\end{equation}
If Joule heating dominates over the particle energy, particle absorption will cause  the TES to undergo only a small shift from equilibrium so parameters in Eqn.\,\ref{eq1} can be linearised. Rather than using absolute temperatures and powers, we can discuss temperature and power changes from equilibrium; without particle absorption, the TES will maintain a temperature of $T_c$ due to Joule heating and thermal conduction losses of $P_0$. With particle absorption,
\begin{equation} \label{a4}
\Delta P_b \approx G\Delta T
\end{equation}

\begin{equation} \label{a5}
\begin{split}
\Delta P_v  &\approx \frac{-V^2}{R^2} \frac{\partial R}{\partial T}
\Delta T \\
            &\approx \frac{-\alpha P_0}{T_c}\Delta T
\end{split}
\end{equation}
where the thermal conductivity, $G$, is defined as $G = \frac{dP_0}{dt}=nKT_c^{n-1}$. $\alpha$ is as defined by Eqn.\, \ref{eq3}. Substituting equations \ref{a1}, \ref{a4}, \ref{a5} into Eqn.\,\ref{eq1} gives
\begin{equation} \label{a6}
C\frac{d\Delta T}{dt} =  -\left(\frac{\alpha P_0}{T_c} + G\right)\Delta T + E_0 \delta(t-t_0)
\end{equation}
which can be solved analytically. The solution takes the form of an exponential decay,
\begin{equation} \label{a7}
\Delta T(t) = \frac{E_0}{C}\exp\left({-\frac{t-t_0}{\tau_{\text{eff}}}}\right),
\end{equation}
with a decay constant of
\begin{equation} \label{a8}
\tau_{\text{eff}} = \frac{1}{\left(\frac{\alpha P_0}{CT_c}+G/C\right)}.
\end{equation}
The decay constant can be simplified under the conditions that $T_c>>T_b$ and $\alpha>>n$ to provide
\begin{equation} \label{a9}
\tau_{\text{eff}} = \frac{nC}{\alpha G}.
\end{equation}
Eqn.\,\ref{a7} describes the temperature dependence of the TES rather than the measured current dependence. Using the relation that $\Delta I \approx -\frac{V}{R^2}\frac{\partial R}{\partial T}\Delta T$, Eqn.\,\ref{a7} becomes
\begin{equation} \label{a10}
\Delta I(t) = -\frac{E_{0}}{C}\frac{\alpha I}{T_c}\exp\left({-\frac{t-t_0}{\tau_{\text{eff}}}}\right),
\end{equation}
as provided in Eqn.\, \ref{eq2}.

\bibliographystyle{aipnum4-1}
\bibliography{ref2}

\end{document}